\documentstyle[12pt]{article}
\begin{document}

\newcommand{\nd}[1]{/\hspace{-0.6em} #1}
\begin{titlepage}
\begin{flushright}
ACT-45\\
CERN-TH.6229/91 \\
CTP-TAMU-66/91\\
\end{flushright}
\begin{centering}
\vspace{.1in}
{\large {\bf On the connection between Quantum Mechanics \\
and the geometry of two-dimensional strings }} \\
\vspace{.4in}
{\bf John Ellis} and {\bf N.E. Mavromatos}\\
\vspace{.05in}
Theory Division, CERN, CH-1211, Geneva 23, Switzerland. \\
and \\

\vspace{.05in}
{\bf D.V. Nanopoulos}\\

\vspace{.05in}
Center for Theoretical Physics, Dept. of Physics, \\
Texas A \& M University, College Station, TX 77843-4242, USA \\
\vspace{.03in}
and \\
\vspace{.05in}
Astroparticle Physics Group \\
Houston Advanced Research Center (HARC),\\
The Woodlands, TX 77381, USA\\
\vspace{.03in}
and \\
Theory Division, CERN, CH-1211, Geneva 23, Switzerland.\\
\vspace{.1in}
{\bf Abstract} \\
\vspace{.05in}
\end{centering}
{\small
\paragraph{}
On the basis of an area-preserving symmetry in the phase
space of a one-dimensional matrix model - believed to
describe
two-dimensional string theory in a black-hole background
which also allows for space-time foam - we
give a geometric interpretation of the
fact that
two-dimensional stringy
black holes  are consistent with conventional quantum mechanics
due to the infinite gauged
`W-hair' property that characterises them.}
\par
\vspace{0.2in}
\begin{flushleft}
ACT-45 \\
CERN-TH.6229/91 \\
CTP-TAMU-66/91 \\
September 1991 \\
\end{flushleft}
\end{titlepage}
\paragraph{}
\newpage
\section{Introduction and Summary}
\paragraph{}
In a recent paper \cite{Ell} we identified the mechanism
by which we believe that quantum mechanics is reconciled
with general relativity, in the context of string theory
in a two-dimensional target space-time. This identification
was based on studies of $c=1$ matrix models \cite{Gro} that
solve, in a certain limit, two-dimensional quantum
gravity
with matter
coupled to a Liouville mode. The latter
can be seen to correspond
to coset $\frac{SU(1,1)}{U(1)}$ Wess-Zumino models \cite{Ant,Witt}.
The crucial observation, due to Witten \cite{Witt}, is that such
models can be interpreted as two-dimensional target space
stringy $\sigma$-models in a
black hole background metric.
It is then
argued that the $c=1$ matrix model is the last stage of the black
hole evaporation.
\paragraph{}
The $c=1$ matrix model can be represented as a theory of
essentially
{\it free fermions},
and as such it possesses an infinity of
conserved currents of high spin \cite{Kleb}. After bosonisation
\cite{Jev} such conservation laws have been seen to
give rise to an infinity of (target space)
conserved charges \cite{Ava},
that are known to form a Cartan subalgebra of a
$W_{\infty}$-algebra \footnote{To avoid confusion
we should mention that the $W_{\infty}$-symmetries we are
talking about are the $w_{\infty}$-symmetries in the notation of
Pope {\it et al.} \cite{Pop}, which admit
central extension only in
the usual Virasoro sector, generated by spin-2 fields. In this sense
the algebra may be considered as
the classical limit of the centrally
extended (beyond Virasoro sectors)
$W_{\infty}$-algebras discovered by Pope, Romans and
Shen \cite{Pop}. It should be noted that
at present a geometric interpretation exists
only for $w_{\infty}$-algebras, as being a
subgroup of area-preserving diffeomorphisms on a
certain two-manifold \cite{Bak}.
It is mainly this
kind of extended conformal algebras that we
shall be concerned with
in this work, and from now on we stick to the $w$-notation,
although in the relevant literature usually there is no
notational distinction \cite{Ava}.}\cite{Pop}.
\paragraph{}
In singular space-times,
symmetries do not imply the existence of
conserved quantities \cite{Hag,Gross}, unless the symmetries
are {\it gauged} by coupling the corresponding currents to long-range
fields \cite{Lahi}. In ref. \cite{Ell} we argued that this is just what
happens in the case of two-dimensional black holes, thus
ensuring the
maintenance of quantum coherence, which in local point-like
theories
has been argued to
be violated due to black hole evaporation \cite{Hawk}.
In other words we identified
the infinite set of hair, associated with
a $w_{\infty}$-algebra of the target space,
which we call `W-hair',
that
reconciles quantum mechanics
with non-perturbative quantum gravity as described by
two-dimensional string theory.
We illustrated this identification in ref. \cite{Ell} by
exhibiting the conserved current that couples to one of the
`massive' topological `Q-graviton' states that constitutes the
last phase of black hole evaporation. Such a discrete string state
is the first of an infinite tower of
stringy states with standard ghost number and
definite
energy and momentum, which in two target space-time
dimensions
constitute the only remnants of the higher-spin string modes \cite{Pol}.
It should be remembered that the `tachyon', which is actually massless
in this case, is the only fully-fledged field of the theory.
We conjectured that all of these massive
modes are associated with stringy
{\it gauge}
symmetries in target space, which lead to the infinity of
conserved charges
of the matrix model, viewed as the last
stage of black hole evaporation
which respects the conservation laws.
\paragraph{}
Subsequently, and independently, Moore and Seiberg
\cite{MS} confirmed the
association of the infinite set of conserved charges in the $c=1$
system with the discrete topological modes of the two-dimensional
string,
by showing that for the particular values of energy and momentum
that characterised the discrete high-spin modes one can find field
transformations that leave the action of the matrix model (in the fermion
formalism) {\it invariant}. They showed that these transformations
generate $w_{\infty+1}$-symmetries for the target space-time.
In particular they found a larger infinite set of conserved charges,
as compared to the results of \cite{Ava}, which obey the
commutation
relations of a $w_{\infty+1}$-algebra. The charges of \cite{Ava}
constitute
a commuting subset (Cartan subalgebra)
of this algebra. In this way Moore and Seiberg
confirmed the existence of
`W-hair'
characterising the black hole solution of the associated
$\sigma$-model theory.
For completeness
we review their results in the fermion
formalism of the $c=1$ matrix model.
\paragraph{}
The action takes
the form
\begin{equation}
     S= \frac{1}{4}\int dt[(\frac{d\psi(t)}{dt})^2 -
 \psi(t)^2]
\label{act}
\end{equation}

\noindent where $t$ is a {\it Euclidean time} and will be
identified with the target space
time in the effective field theory (after a Wick rotation).
Symmetries of the action are found by redefining
the fermion variables
 $\psi(t)$ according to \cite{MS},

\begin{equation}
 \delta^{s, l} \psi(t)= i\frac{\partial}{\partial p}[H^{s, l}]
e^{ilt}
\label{var}
\end{equation}

\noindent where $p=\frac{d\psi(t)}{dt}$,
$s$ is a positive integer, $l=s,s-2,...-s$,
and $H^{s, l}=(-ip-\psi)^{\frac{s+l}{2}}
(-ip+\psi)^{\frac{s-l}{2}}$. For these particular
values of $l$ the variation of the action (\ref{act})
vanishes. It can be shown from the analysis of \cite{MS}
that these values precisely correspond to the extra discrete
modes that we mentioned above. Such modes may be seen to
correspond formally to what one would call `boundary operators'
in matrix models \cite{MS2},
which however for {\it the particular}
values of momentum $l$ given  above become `bulk' operators
corresponding to the discrete massive high-spin states
of the two-dimensional string theory obtained as a certain
continuum
limit
of the matrix model.
This infinite
family of
symmetries is generated by conserved charges of the
form $O^{s, l}=H^{s, l} e^{ilt}$  which are shown to
obey the commutation relation of a $w_{\infty +1}$ algebra,
\begin{equation}
      [ O^{s, l}, O^{s', l'}]=(s'l - sl')O^{s+s'-2, l+l'}
\label{alg}
\end{equation}

\noindent The charges discussed in \cite{Ava}
the beginning
correspond to the
commuting subgroup $l=0$.
Notice the explicit time dependence
of the charges for $l \ne 0$. This is analogous to the case
of the generators of Lorentz boosts in conventional field theories
\footnote{We thank L. Alvarez-Gaum\'e for an illuminating discussion
on this point.}.
In view of the conjectured equivalence
of the
matrix model and $\sigma$-model formalism
one
expects that upon bosonising the theory, by taking into
account
the extra operators,
one would recover
this set of symmetries as target space-time symmetries
of the corresponding
(1+1)-dimensional $\sigma$-model. In this way the conserved
charges could be expressed as functionals
of the target-space
backgrounds integrated over the `space' (Liouville coordinate)
of the two-dimensional string \cite{Ell}. In this framework
the symmetries of the  matrix model could be viewed as
{\it exact} (anomaly-free) target-space
{\it gauge} symmetries of the $\sigma$-model, formulated
in higher genera.
\paragraph{}
There are two ways in which
such symmetries are realised in the $\sigma$-model
approach to string theory.
In one of them,
the symmetries
correspond to
redefinitions, $\delta X$,
of the $\sigma$-model target space coordinates $X$
that leave the Fradkin-Tseytlin (FT)
functional invariant \cite{Ven}.
In this way the symmetries are dissociated (at least formally)
from the conformally invariant critical point, given that
the FT generating functional can be defined away from the
fixed point \cite{Zam}.
In the alternative approach \cite{Ovr},
conformal invariance
of the underlying world-sheet
field theory is essential.
In this approach the
symmetries are associated with a
family of
string vacua.  A target space stringy symmetry in
this case
would imply the existence
of infinitesimal (or finite) conformal field theory operators
that generate isomorphisms between conformal field theories
with target space fields $\{ g^i \}$ and $\{ g^i + \delta g^i \}$
\cite{Ovr}.  The formal criteria for such a symmetry
can be summarised as follows \cite{Ovr},
\paragraph{}
(i) Any infinitesimal
world-sheet operator $h$ generates
an isomorphism
between conformal field theories with stress tensors
$T$ and $T+i[h,T]$ \footnote{We use
Euclidean notation with
$T_{zz}\equiv T,
T_{{\bar z}{\bar z}}\equiv {\overline T}$,
and $T$ is the total stress tensor including the ghost
contributions. It is related to the Hamiltonian $H_g$
of the $\sigma$-model by $\int d\sigma (T + {\overline T})$,
with $\sigma$ a spatial world-sheet coordinate.}.
\paragraph{}
(ii) Two conformal field theories with
stress tensors that differ by an operator $\chi_{(1,1)}$
of conformal
spin (1,1) correspond to backgrounds that differ
by an infinitesimal change appropriate to $\chi_{(1,1)}$.
This follows from the assumption that in a $\sigma$-model
language one considers as physical (background)
space-time fields
the ones that are in one-to-one correspondence
with primary fields of dimension $(1,1)$. However
this restriction does not prevent other operators
of the theory from being important in
determining
the structure of space-time symmetries, as we
shall discuss shortly.
\paragraph{}
(iii) Symmetries of the target space of the string
are, therefore, generated by currents $J$
of conformal
spin (1,0) or (0,1) (this is so because in that case
the charge $\int d\sigma J \equiv h $ has the property
that $i[h,T]=\chi_{(1,1)}$ \cite{Ovr}).
\paragraph{}
It is understood that appropriate extensions
to finite transformations are expected to hold,
but the situation is not clear. Discrete string
symmetries might form a different category,
although the belief is that they also can be
incorporated in the second scheme.
\paragraph{}
It is not clear at present whether the two
approaches are equivalent. Certainly there are examples
where a change of variables in the FT functional induces
non-trivial changes in the background fields, but it is
not yet known whether this includes the full set
of stringy symmetries \footnote{For Bosonic strings,
the higher gauge symmetries corresponding
to
massive multiplets up to
rank four
are known to
be obtained from appropriate transformations of the
target-space
coordinates $X^{M} \rightarrow X^{M} + \zeta^{M}_{NP}(X)
\partial X^{N} {\overline \partial} X^{P}, \partial$ being
a world-sheet derivative \cite{Mav}.}.
We conjecture \cite{Ell}
that
in the particular case of two-dimensional strings, the
$w_{\infty+1}$-symmetry of the target space can be
represented as a change in $\sigma$-model fields.
We base this conjecture
mainly on intuition at this stage,
but also
on some preliminary
results in simplified systems characterised
by $W_{\infty}$-symmetries \footnote{We thank E. Kiritsis
and I. Bakas for interesting discussions on their preliminary
results on this issue.}. Our intuition is based
on the so-called $W$-gravity systems, which are known
to possess extended conformal symmetries \cite{W-grav}.
For instance,
the action
of a two-dimensional
free
complex scalar field, $\int d^2 z
\partial \phi {\overline \partial} \phi^{*}$,
is known to possess
$W_{\infty}$ as well $w_{\infty}$-symmetries,
which are represented
as appropriate changes of the field $\phi$ \cite{W-grav}.
A complex scalar field may be considered as
parametrising a
complexified two-dimensional target
space, and the above theory is nothing other
than string theory in a flat space-time background
\footnote{It goes without saying that this
correspondence is only formal since string theory
on flat backgrounds cannot be formulated
consistently in two target-space
dimensions. To this end one should employ
non-trivial backgrounds.}.
In this picture the world-sheet symmetry is realised
as a transformation of the spacetime coordinates \cite{Ven}
which
leaves the flat background invariant, and in this sense
it is trivial from a target
space-time point of view
\footnote{There is an important difference from the usual
critical string case, in that the target space coordinate
transformations
contain an explicit dependence on the
world-sheet coordinates which appears in the respective
parameter. For example the flat background
$W_{\infty}$ symmetry corresponds to changes of the
field $\phi$ of the form \cite{W-grav}
$\delta_{k_{l}}
 \phi =
\sum_{k=0}^{l} \alpha^{i}_{k}
\partial^{k} (k_l \partial^{l-k+1} \phi)$,
where
the parameters
$k_l$ are {\it holomorphic}
functions of the world-sheet
coordinate $z$. It is interesting to
note \cite{Shen} that by `fermionising'
the complex boson problem, i.e. by
defining a pair of fermions $\eta$ and $\chi$
such that $\phi \equiv  :\eta e^{-\sigma} :$ and
$\phi^{*} \equiv :\partial \chi e^{\sigma}:$
(where $:...:$
denotes the
appropriate normal ordering),
one can
obtain non-linear realisations
of
$W_{\infty}$ and $W_{\infty+1}$.
By further bosonisation of fermions \cite{Shen}
one can then
obtain - after appropriate truncation -
realisations of the
{\it classical}
$w_{\infty}$-symmetries
in terms of two real scalars, which may be
(formally) closer to the stringy
target-space interpretation
argued above.}. Once non-trivial
backgrounds (field interactions)
are introduced
one expects that
generalised coordinate transformations  will induce
changes in these backgrounds, and in this way
a world-sheet symmetry is elevated into a target
space one. We argue below that this is
precisely the case with the particular
background corresponding to matrix models.
\paragraph{}
In any case, irrespectively of the form  of the symmetries
in the $\sigma$-model language, their existence
would lead to Ward identities
of amplitudes
mixing various string levels \cite{Ven,Ell}. They would correspond
to higher-spin massive string
states
\footnote{In view of this mixing
of levels of different mass one would expect that in higher-dimensional
string theories such symmetries would be spontaneously broken
by the background. From the analysis of \cite{Ovr} it is evident
that the order parameter
of such a breaking is the metric for space-time. An interesting
result is obtained by noting that the symmetries are not
restored
for any value of this field. This has been associated
in \cite{Ovr} with the symmetric phase of the string where
an S-matrix cannot be constructed \cite{Witten}.
In the particular case of
two-dimensional string, spontaneous breaking cannot
occur due to Coleman's theorem \cite{Com}. In contrast,
one might encounter a Kosterlitz-Thouless type of symmetry
breaking. The topological nature of the space-time graviton
mode in this case \cite{Pol}, and the fact that such excitations
characterise the final stage of black hole evaporation
\cite{Ell,Mand},
might then prove useful in understanding
the nature of the
string vacuum characterising the matrix model system.}.
 The existence of such
operators and the corresponding Ward identities
in the  matrix model \cite{MS}
certainly points
in this direction,
but more work needs to be done, especially in higher genera, before
rigorous results are derived.
\paragraph{}
Recently Witten \cite{Witt2}
has managed to exhibit a larger symmetry for the $c=1$ matrix model
by making use of new  discrete states
of non-standard ghost number, which were known
in the
previous literature \cite{Lian},
but not extensively discussed.
The basic feature is the existence
of modes of ghost number zero and conformal spin zero
that form a ring, the
so-called `ground ring'.
In terms of matrix models this ring helps in
understanding the
symmetry of the system,
which is nothing other
than a phase-space area-preserving
transformation that leaves the Fermi surface invariant.
It is important that the world-sheet currents that
generate this particular Lie algebra of symmetries
are operators of non-standard ghost number but of
conformal spin $(1,0)$ and/or $(0,1)$.
Given that in the discussion we gave above on `hidden'
symmetries in string theory
the ghost number seems to play no essential role, we have
an understanding of this symmetry as an ordinary stringy
symmetry of target space-time of a two-dimensional
$\sigma$-model. The fact that the symmetry is known in a closed
form is due to the low
dimensionality of target space.
The
symmetry is bigger than
that discussed previously
\cite{Ava,Ell,MS},
but
the extra symmetries are quite trivial as far as the
matrix model is concerned. However,
they are important
in the conformal field theory framework.
As we mentioned in the beginning of the article,
it has been known for some time \cite{Bak}
that $w$-algebras
are
associated with specific
area-preserving diffeomorphisms for fields of
high spin,
but
the connection with string theory had not been clear.
\paragraph{}
In the body of this note we develop a physical interpretation of
this remarkable observation in the context of the
symplectic
geometry of Hamiltonian dynamics. Specifically, it ensures that
the matrix model's
time evolution respects Liouville's theorem,
which is a necessary
condition for canonical quantisation. Comparing with
a generalised evolution equation for the density matrix
proposed previously \cite{Hag} to accommodate the breakdown of quantum
coherence conjectured by Hawking \cite{Hawk}, we demonstrate
the vanishing of the extra term that otherwise
would allow
pure states to evolve into mixed ones.
The basic feature is that the extra terms violate the exact
stringy symmetry of the model,
which is expressed as the transformation preserving the
phase-space volume element
\footnote{It should be stressed that this symmetry is a
property of two-dimensional {\it strings}, associated
with the infinity of modes characterising
various string multiplets, which
does not exist in
conventional local field theories.}.
This
geometric view-point on the
restoration of quantum
coherence in two-dimensional strings
probably opens new ways to attack
similar problems
in higher-dimensional string
theories.
We also recall that
the area-preserving diffeomorphims of
the matrix model leave invariant the Fermi surface \cite{Witt2}.
An analogous result in four-dimensional field theory would
reflect the existence of a large string symmetry interelating
vacua with the same value of the cosmological constant $\Lambda$.
Assuming the existence of a supersymmetric vacuum with zero
cosmological constant, $\Lambda=0$, the spontaneous
breakdown of this large string symmetry would
leave string in some asymmetric vacuum with $\Lambda=0$.
\paragraph{}
\section{The matrix model, black holes and space-time foam}
However, there is a non-trivial conjecture
that one has to make about the nature of the matrix model
in order to apply the
above arguments. The conjecture concerns
the role of the black hole as well as
space-time foam in this context. The matrix
model is believed to be the last stage of black hole evaporation
\cite{Witt}.
Black hole solutions have been
discussed so far in the
context of solving $\beta$-function
conditions for
$\sigma$-models
formulated on genus zero,
while the matrix
model is considered as a stringy regularisation of world-sheet
quantum gravity where the summation over genera is done exactly.
It is quite plausible that higher genus corrections do not
destroy the black hole character of the
tree level conditions, and this will be assumed in the following.
The black hole solution is asymptotic (at spatial infinity)
to a space-time which is exactly that of a matrix model.
The symmetries of the matrix model in this way become
asymptotic symmetries of space-time {\it in the presence
of a black hole} that can be assumed to evaporate somehow.
Given that under time evolution the phase space of the
matrix model, which is essentially the phase space of
$c=1$ matter interacting with the black hole, is preserved
as a result of the stringy symmetry, one has an elegant
geometric interpretation of the maintenance of quantum coherence
in the system.
\paragraph{}
Is that the end of the story? In other words, is the classical
black hole background and its evaporation the only source of loss
of coherence in the system?
One might naively answer positively, given that after the
black hole evaporates one recovers flat space-time
and the
problem of defining an S-matrix that factorises in the case
of matrix model seems quite natural and somehow trivial.
However,
the essential point that will concern us below lies in the fact that,
if one takes the dynamics of space-time seriously in the
case of $c=1$ theories, there are additional complications
even after the black hole evaporation, that can be seen on a
{\it microscopic} scale. The problems come from the
observation that a background-independent formalism
of the Das-Jevicki action \cite{Jev} is possible,
which
implies that if
one considers
quantum fluctuations around any background, then
creation and annihilation of {\it virtual}
black holes may occur, and as a result
space-time foam
will appear.
\paragraph{}
Let us be more precise.
First we
argue how
one can formally achieve a background-
independent formulation.
 From the collective field theory approach to the matrix model
\cite{Jev,Sak}, it becomes evident
the latter
yields a {\it string field theory} of tachyons
that propagate in a {\it flat}
space-time configuration,
where the dilaton field is linear in the
the spatial coordinate and
all the rest of the higher-spin string modes are vanishing.
The reason for having a tachyon field theory and not simply
a classical background
is simple. One starts from the {\it partition function}
of the lattice model and then makes a change of variables
(which is a
{\it canonical} transformation that leaves invariant
the classical phase-space volume element)
following \cite{Sak}
to pass
to a target space field formalism.
In this way, the collective field coordinate is integrated
over in the path integral. This coordinate can be seen to
correspond to the tachyon field \cite{Jev,Polch2}.
There is no dynamics in the matrix model for the high-spin
string modes, by construction.
 From classical
symmetry considerations \cite{Ava}, however, one
can anticipate a background-independent formalism, which
will enable one to integrate over the
space-time fields and obtain a {\it quantum}
string field theory.
To this end,
one has first to
covariantise the Das-Jevicki action. That this is
possible follows directly from the fact that
the amplitudes of the Das-Jevicki theory are nothing other
than string theory amplitudes for the tachyon field
in
a two-dimensional target space
\cite{Pol}. The insertion of
graviton-dilaton states with
generic momenta in tachyonic amplitudes
can be seen to
yield vanishing results, which can be interpeted \cite{Ell}
as arising from certain stringy Ward identities expressing
general covariance of the theory.
In this way general
covariance
is viewed as one of the
the infinite gauge symmetries of the two-dimensional string
associated with massive high-spin levels \cite{Ven}.
It is straightforward \cite{Ell},
following the dictionary outlined in \cite{Jev,Polch2},
to express the Das-Jevicki lagrangian in terms of
a tachyon field propagating in non-trivial (generic)
graviton and dilaton backgrounds. The result is
an
effective action, involving an infinite number
of derivatives,
 describing the interaction of
the tachyon with these fields. The infinite derivative order
is immediately
seen from the fact that the Das-Jevicki lagrangian \cite{Jev}
for the collective field $\phi(x,t)$
contains,
in the scaling limit,
kinetic terms of the form
\begin{equation}
 \int dx \frac{1}{2}[\frac{(\partial_x^{-1}
  \frac{\partial \phi}{\partial t})^2 }{\phi} + L(\phi)  + O(\phi^3)]
\label{DJL}
\end{equation}

\noindent where the interaction terms
$L(\phi)$
depend on the details
of the matrix model potential.
For a
study of
the critical properties
of the model it
suffices to consider
inverted harmonic oscillator potentials
\cite{Jev,Polch2}. In this case, $L(\phi)=(\mu_F+\frac{1}{2}x^2)\phi$,
where $\mu_F$ is the chemical potential of the matrix model \cite{Jev}.
It should
be noted that
$\mu_F$ is essential for the
non-perturbative
(field-theoretic)
quantum corrections we are interested in,
although it may be ignored when one considers only
classical tachyon backgrounds
\cite{Polch2}.
To get a covariant form,
one may redefine the collective field by looking at
fluctuations around static
classical minima, $v_0(x)=\frac{1}{\pi}\sqrt{2\mu_F+x^2}$,
of the effective potential
corresponding to
(\ref{DJL}). Thus we write
$\phi =v_0(x) +  \frac{1}{v_0(x)}\eta(x,t)$,
change
variables from $x$ to a space-like coordinate  $q$
(the Liouville mode
\footnote{It should be noted at this stage that it is not
at all obvious that we can establish an exact (formal)
correspondence of the spatial coordinate $q$
appearing in the Das-Jevicki action with the Liouville mode
of the continuum model. In view of the results of \cite{MS}
this appears to be true only for low energies, while in general
there is a non-local integral transform that connects the two
parameters, which makes the connection of the two theories
technically obscured. Indeed, it can be readily seen
that in the low-energy
limit
the equations of motion obtained from the
action (\ref{nl})
are equivalent to the
lowest order ($\alpha'$) $\sigma$-model $\beta$-function
conditions. However the situation is not at all clear
with the higher-order corrections, since exact
results (on higher genera) from the
$\sigma$-model are not yet available. One should hope
that when taking into account the field redefinition
ambiguities in the $\sigma$-model formulation and
the relevant subtleties \cite{MS}
of the connection
of the Liouville mode and the spatial coordinate of the
Das-Jevicki action, the connection
of the two formalisms might
be elucidated.
 This immediately raises the question of
which theory
is more convenient (or even correct)
 to describe the
dynamics of the
$c=1$ theory.
The point of view we took in \cite{Ell},
and which we follow here, is that the stringy
approach
is more fundamental. After all,
it is essentially this approach
that gave rise to black-hole solutions \cite{Witt,Mand}.
The connection of the conformal field theory with the matrix
model, and therefore
the Das-Jevicki formalism,
is then elucidated further in \cite{Witt2}
by exploiting the extra structure of
the conformal field theory,
that cannot be seen in the matrix model approach. It is
probably this extra structure \cite{Min}
that explains a discrepancy
between the works of Gross and Klebanov \cite{Tri}, on the
triviality of the S-matrix of $c=1$ matrix model theory, and that of
Polyakov \cite{Pol}.
It is in view
of this connection that the background-independent formalism
of the Das-Jevicki string field theory is useful.})
defined by, $q=\int ^{x} \frac{dx}{v_0}$,
and define
a tachyon field $T$ by $\eta \equiv \partial_q T(t,q)$.
One
immediately observes
that the kinetic terms of (\ref{DJL}) when
{\it combined} with the potential terms $L(\phi)=
\frac{1}{\phi}L(\phi^2)$
yield (after a Wick rotation in time $t$)
terms of the form
\begin{equation}
           \int dq dt \frac{(\partial_t T)^2 -
 (\partial_q T)^2 + O[v_0(q;\mu_F),\partial_q T(t,q)]}
{1+ \frac{1}{v_0^2(q;\mu_F)}
 \partial_q T}
\label{nl}
\end{equation}

\noindent where for the shake of simplicity we denoted
collectively the interaction terms by $O[...]$.
It is immediately seen that that the covariantisation
is obtained naively by setting $2q \rightarrow \Phi$
(a dilaton field),
$\partial_q T \rightarrow \frac{1}{2}
 G_{\mu\nu}\nabla_{\mu} \Phi \nabla_{\nu} T $, and
$ (\partial_t T)^2 - (\partial_q T)^2 \rightarrow \nabla_\mu T
\nabla_\nu T G^{\mu\nu}$. The standard $\sigma$-model
form is obtained by redefining $T'=e^{-\Phi}T$, which induces
the non trivial dilaton factors in the integration measure
\cite{Witt,Ell}
$\int d^2y e^{-\Phi} \sqrt{G} ...$.
There are ambiguities in
such a covariantisation since the matrix model is formulated
on a {\it flat} target
space-time and therefore
 curvature terms, which determine
space-time dynamics, are vanishing. Moreover,
one cannot
really distinguish between constant factors in the action and
purely dilaton terms.
 To resolve such questions one has to
to start from a stringy point of view \cite{Ell,Pol}.
If one is interested only in classical space-time dynamics
for the high-spin modes (which in two dimensions have definite energy
and momentum),
the
$\sigma$-model point of view seems sufficient.
The latter
can be re-expressed, at least
as far as the
graviton-dilaton multiplet is concerned
\cite{Witt},
as a $\frac{SU(1,1)}{U(1)}$
Wess-Zumino model. However,
to get a consistent string theory
one should also
include the
rest of the string modes that couple to the tachyons \cite{Ell}.
That such couplings exist becomes clear from the $\sigma$-model
formalism, where the stringy gauge symmetry transformations
generated by
these modes mix the various string levels \cite{Ven,Ovr,Ell}.
These symmetries exist already at the conformal field theory
level, i.e. a first-quantised string formulated on
flat space-times, imposing non-trivial constraints
(Ward identities) on string
S-matrix elements \cite{Ven,Ell}. Such constraints
can be interpreted as arising from symmetries
of the target space-time associated with the rest
of the spin modes that are vanishing in the
fixed background in which
we are working. To find the precise
form of the relevant terms is formally a
complicated task,
and at present there is
no closed form known for the collective field theory
to incorporate these modes, but in principle it exists.
Nevertheless,
following the above
methods of covariantisation,
one can
show  explicitly
\cite{Ell}
that the first two lowest-lying
conserved charges
of the effective theory associated with the string levels
zero and one
(fermion number and the spin-two current
related to the energy of the black hole)
are surface terms,
and as such
characterise
the
asymptotic black hole state  and are, therefore,
quite insensitive to what is
going on inside the horizon. One can anticipate \cite{Ell}
a similar result for the {\it entire} set  of the
conserved charges of the matrix model, associated with
other string modes.
In this way one has an infinity
of independent
charges that are
exactly conserved even during the
Hawking process of black hole evaporation. They
constitute
the gauged `W-hair' of the black hole \cite{Ell},
which we
argued to be
responsible for the
restoration of
quantum coherence, since
they are remnants of the black hole after the
evaporation.
\paragraph{}
When gravity
and higher-spin fields
are included, hopefully
through
an appropriate
extension of the Das-Jevicki approach \cite{Jev},
one should integrate over
all possible configurations
of these fields. Although
these modes are discrete string
states of definite energy and momentum, as can be readily seen
in an expansion
around  {\it flat} target
space-time
\cite{Pol,Ell},
in general
they may give rise
to various field
configurations,
which can only be seen if one starts
from a background-independent
string field theory and not expanding around a {\it given} background,
as in the $\sigma$-model approach.
In such a procedure one should
include
topology-changing configurations (in Euclidean formulations),
which are in any case included in the set of allowed classical
solutions of the $\sigma$-model $\beta$-function conditions.
It is therefore natural to assume
that the non-perturbative solution to the matrix model includes
the creation and annihilation of virtual {\it microscopic}
black holes.
In this interpretation Witten's black hole, which is a
solution of the beta function conditions of the $\sigma$-model
at genus zero,
plays the role of
what Hawking would call a {\it macroscopic}
black hole \cite{Hawk}.
The $\sigma$-model (formulated in higher genera)
would involve quantum
fluctuations of the space-time metric
which lead to
the formation of virtual black holes and microscopic horizons.
The latter
survive
the evaporation of the macroscopic black hole,
and thus characterise the asymptotic quantum field theory,
which in this case may be
interpreted as
describing
the
interaction of $c=1$ matter with a space-time foam whose role is
similar to that of
a quantum-mechanical reservoir.
\paragraph{}
Hence,
in principle,
the modification of the
quantum mechanical evolution suggested in \cite{Hag}
could apply.
However,
this might not be the case due to stringy symmetries, as
we shall argue below.
There is also a nice
geometric intepretation
of these
symmetries,
which also illuminates
the relation between matrix model and string theory in fixed
backgrounds.
\paragraph{}
\section{The symplectic geometry of Quantum Coherence}
\paragraph{}
We now proceed
to demonstrate the geometric origin of the argued
restoration
of quantum coherence in the two-dimensional string.
   To understand better the problem and its solution in the matrix model
framework
it is first useful to recall some  facts about the
possible modifications of quantum
mechanics as a result of singular space-time metrics \cite{Hag}.
\paragraph{}
Consider a quantum mechanical problem described by
a Hamiltonian $H(p,q)$ and a density matrix $\rho(p,q,t)$.
In conventional quantum mechanics,
the time evolution
of the density matrix is given by
\begin{equation}
\frac{\partial \rho}{\partial t}=i[\rho, H]
\label{dens}
\end{equation}

\noindent where $[,]$ denotes the
conventional commutator.
Eq. (\ref{dens}) is nothing other
than the quantum-mechanical
analogue of Liouville's theorem for the conservation
of the classical probability density under time evolution
in classical
mechanics.
There are two features that lead to such
a classical
conservation.  For our purposes it
is instructive to
review them
briefly.
\paragraph{}
The {\it first}
is the fact that the subsequent motion of the
classical system (defined by a point in phase space) is
completely determined by its location in phase space at a
particular time. This implies that the number of systems
inside the infinitesimal volume element surrounding a given point
remains constant under time evolution. A system initially inside
the volume can never get out,
since at
the moment of crossing the
border
it has
the same location as one of the boundary points and therefore
both points would be forced to travel together from that moment.
The {\it second}
important point is that time evolution is a canonical
transformation,
and as such leaves the infinitesimal volume element
in phase space invariant.
Thus,
the density of phase-space points remains constant during
time evolution.
\paragraph{}
Eq. (\ref{dens}) might be modified if singular space-time
metric backgrounds were taken into account, due to the
Hawking evaporation
phenomena that accompany them \cite{Hag}. Already in ref. \cite{Hawk}
Hawking had proposed that once non-perturbative metric
fluctuations were taken into account, the usual S-matrix
relation between initial and final states should be replaced
by a general linear relation
\begin{equation}
              \rho_{out}=\nd{S} \rho_{in}
\label{smat}
\end{equation}

\noindent where $\nd{S}$ cannot in general be factorised
into the product of $S$ and $S^{\dagger}$, as in conventional
quantum field theory. It was suggested in ref. \cite{Hag}
that the corresponding modification of eq. (\ref{dens})
would be
\begin{equation}
  \frac{\partial \rho}{\partial t}=i[\rho, H] +
{\hat{ \nd{H}}}\rho
\label{dens2}
\end{equation}

\noindent where the extra part represents an abstract operator,
whose typical effects are suppressed by powers
of the Planck scale \cite{Hag}.
The immediate
interpretation of this modified
quantum mechanics would be that
gravity induced non-invariances
of the
phase space volume of the system.
The existence of microscopic event horizons
might cause similarities
with the situation in {\it open} systems,
where there is a flow of
information to and/or from the system, and this simply induces
a failure of Liouville's theorem for the open system.
Indeed, eq. (\ref{dens2}) is similar in form to the
markovian master equation that controls the evolution
of a reduced quantum mechanical system in interaction with a
reservoir \cite{Gor}. The important difference \cite{Hag}
is that in this scenario for
quantum gravity
the role of the reservoir
is played by the space-time foam, which however is
intrinsically unobservable.
Therefore the modification
(\ref{dens2}) is essential,
and not an artefact of
a voluntary choice
to observe only
a reduced part of the world.
\paragraph{}
Of the two features stated above that characterise the motion
of an isolated classical Hamiltonian system, the one that is
violated in the case of open systems
is the second, namely the invariance of the phase space
volume element under the time evolution, due to the non-existence
of unitary Hamiltonian motion. This can
be understood  by the fact that in the case of the evolution
of the open system there is flow of information through the
boundaries. This is effectively described by the non-invariance
of the phase-space differential volume.
 As a
 result,
the equation for the
probability density,
and  evidently the density matrix
evolution,
is modified by the extra term ${\hat{\nd{H}}}$
in
(\ref{dens2}),
whose precise form depends on the details
of the interaction of the subsystem in question with the reservoir.
We do not know how such a modification due to space-time foam
would be avoided
within the context of conventional local quantum  field theory.
\paragraph{}
However,
in the  particular case of two-dimensional string
black holes (matrix models)
the conservation of the phase space volume is guaranted in the
quantum system by a {\it symmetry} of the theory,
as we shall now discuss,
and as such the above modification of
quantum mechanics
is not present. This is the promised
geometric interpretation of the
restoration of quantum coherence argued in \cite{Ell} on the basis
of the
infinite gauged `W-hair'
property of the two-dimensional string
black hole.
\paragraph{}
Following Witten \cite{Witt2},
we consider
the matrix model Hamiltonian which has the standard
inverted harmonic oscillator form,

\begin{equation}
   H(p,q)=\frac{1}{2}(p^2 - q^2)
\label{ham}
\end{equation}

\noindent The action
density of the model in form language is given
by the two-form  $\omega =dp\wedge dq - (pdp-qdq)\wedge dt$,
where t is a time parameter. Solving the model means to find
an appropriate change of variables in ($p,q$) space
(as functions of $t$)
so that the equations of motion $\frac{dp}{dt}=-
\frac{\partial H}{\partial q}$ and $\frac{dq}{dt}=\frac{\partial H}
{\partial p}$ look trivial. The transformation is readily obtained
\cite{Witt2}
\begin{eqnarray}
\nonumber p'& = & pcosht-qsinht \\
  q' & = & -psinht + qcosht
\label{trns}
\end{eqnarray}

\noindent Indeed,
the equations of motion in the  $p'-q'$ frame
are $\frac{dp'}{dt}=0$ and $\frac{dq'}{dt}=0$.
In the ($p',q'$)-system of phase space coordinates the
action density itself becomes the phase space volume element
$\omega =dp' \wedge dq'$.  It is therefore evident that the
hamiltonian vector fields, which by definition leave the
$dp'\wedge dq'$ element invariant, are {\it symmetries} of
the matrix model. Such fields are easy to construct \cite{Witt2},
\begin{equation}
       {\bar{V}}=\frac{\partial g(p',q')}{\partial q'}
       \frac{\partial}{\partial p'}-
       \frac{\partial g(p',q')}{\partial p'}\frac{\partial}
      {\partial q'} + u(p',q',t)\frac{\partial}{\partial t}
\label{vec}
\end{equation}

\noindent In terms of the original $p-q$
system,
the above vector fields
generate symmetry transformations for the matrix model.
In particular,
$g(p',q')$
is associated with shifts in the $q$-coordinate,
while $u(p',q',t)$
represents changes in the time coordinate $t$.
It is clear
that the time evolution of the quantum system in the $p'-q'$ plane
becomes in this way an exact symmetry of the conformal
field theory \footnote{Actually the only restriction in that case
is that the function $u$ does not have an explicit dependence
on time.}.
Hence,
the classical result of the invariance of the
phase space volume element
is promoted in this particular case into a symmetry of the problem.
As a consequence, the
Liouville-von Neuman evolution of
the density matrix in ordinary quantum
mechanics is {\it guaranteed} by the
symmetries of the theory.
 From the above-described equivalence
of the matrix model and the Das-Jevicki
tachyon field theory,
one can anticipate that the phase
space of the latter will remain constant under time evolution,
for stringy symmetry reasons.
Witten's analysis \cite{Witt2} shows that
this evolution corresponds to a
symmetry of the conformal field theory which
is associated with
states of given energy and momentum characterised by
non-standard ghost number
adjacent to the one characterising the states discussed in
refs. \cite{Pol,Ell,MS}.
The extension of these symmetries to
non-trivial backgrounds,
which incorporate black holes,
is non-trivial to check, especially in view of the
non-existence, as yet, of a satisfactory string field theory
formalism that resums higher genera in the continuum.
However, in view of
the above-described background independent
formalism of the Das-Jevicki action and its appropriate
extension to
include high-spin string modes,
one should anticipate that such
an extension is
possible.
In this case, taking into account
the above interpretation of the role of space-time
foam in this context,
and the connection of the asymptotic states of the black hole
to
the matrix model,
one has an elegant geometric interpretation of the restoration
of quantum coherence in the two-dimensional string theory.
The basic feature is the existence of a stringy symmetry,
associated with various string excitations,
which
characterises the ground state of the system.
As a result,
the extra term
in (\ref{dens2}) that
would spoil conventional quantum mechanics is
forced
to vanish, as it is incompatible with the symmetries of the
theory.
This happens due to the fact that the symmetry in question
finds a nice interpretation as a phase-space volume-preserving
diffeomorphism which is precisely the  property
violated in
the modified evolution equation (\ref{dens2}). The existence
of the `W-hair' \cite{Ell}, that characterises the asymptotic
states of the black hole (matrix model),
certainly offers field-theoretic
support to these arguments.
\paragraph{}
The area-preserving diffeomorphisms in phase space are a
feature of the $w_{\infty}$ algebras \cite{Bak}
discussed at the beginning of this
 article. It should be stressed
that the $w_{\infty}$-symmetry that characterises the
standard ghost number discrete states of high spin is included
in the above symmetry. As we mentioned above,
from the matrix model point
of view the extra symmetries are trivial and do not lead to extra
conserved charges. One should probably anticipate that the
charges discussed previously,
associated with the standard ghost number string
states, are the only ones that characterise
the conformal field theory and therefore the black hole solutions.
However,
it is the entire spectrum of string states,
including the non-standard ghost number ones, that
are really responsible for the maintenance of
quantum coherence and the
restoration of the relation between symmetries and conservation
laws. {\it
Hence it seems that it
is the particularity of `W-hair',
which is area-preserving and
characterises the two-dimensional black hole,
that is relevant for the reconciliation of quantum
mechanics and general relativity, and that just having
an infinity of conserved charges may not suffice}.
\paragraph{}
In view of this result,
one might speculate
that if the
above scenario is to be carried through in higher dimensional
space-times, it is necessary that
there exist analogous
infinite-dimensional symmetries associated with area-preserving
diffeomorphisms in some phase space variables.
One might need something like the
$w_{\infty}$-algebras that characterise four-dimensional
self-dual Euclidean Einstein spaces \cite{Park}.
There, it is known that four-dimensional
self-dual Einstein gravity
follows as the large N-limit of a two-dimensional
Wess-Zumino SU(N)
$\sigma$-model.
Infinite-dimensional symmetries of
Einstein's equations
are shown to form a $w_{\infty}$-algebra.
The self-duality of the model
may
not be
essential \cite{Park} for
the existence of infinite dimensional integrability,
and one might
hope that non-self-dual four-dimensional
gravity models obtained
from string
theories with higher-dimensional target spaces might possess
extended conformal symmetries of the typed discussed in this note.
\paragraph{}
\section{Remarks on the cosmological constant problem}
\paragraph{}
After these speculative remarks about the
possible extension
of our
ideas to higher-dimensional space-times,
we conclude with some comments on the cosmological constant
in the light
of the above insights into matrix models. Witten \cite{Witt2}
has observed that the infinite set of string symmetries
that preserve the two-dimensional area of phase space of the matrix
model also leave invariant the Fermi surface
\begin{equation}
       F\equiv \{ \frac{1}{2}(p^2 - q^2) \}=0
\label{fermi}
\end{equation}

\noindent If this property of zero vacuum energy
also could be lifted to realistic
string models with an effective
field theory in four-dimensional target space-time,
it could explain why the cosmological constant $\Lambda=0$.
We presume that such theories would have a large number
of consistent vacua, among which would be a supersymmetric
one with $\Lambda=0$. As in the case of a
conventional finite-dimensional field theoretical symmetry,
this vacuum might be invariant under the infinite-dimensional
string symmetry
\footnote{We note parenthetically that
the two-dimensional cosmological constant operator is central within
the string symmetry algebra.}.
However,
there is an
equally generic possibility that this is just one of a degenerate
set of string vacua which are transformed into each other
by the string W-symmetries.
The theory must choose between these
by breaking
the string W-symmetry
spontaneously, analogously to the Goldstone or Higgs
mechanism in local field theory. A key role is played in this
idea by supersymmetry, which ensures that the set of $\Lambda=0$
vacua is at least non-empty. If the set is indeed degenerate,
this should be apparent in the effective four-dimensional field
theory,
which
could well exhibit a set of candidate vacua that have
`small' values of the cosmological constant, at least in some
approximation. Relevant examples of such effective field theories
are the flat-potential no-scale supergravity models \cite{Sup},
which have
a continuum of degenerate
tree level vacua that break supersymmetry, and appear naturally
in string theories \cite{Sustri}. It would
indeed be
remarkable if the two fundamental
problems of the
restoration of
quantum coherence and the
vanishing of the
cosmological constant could be solved within the framework
of a single string symmetry.
\paragraph{}
The relation of these ideas to those of \cite{Col,Ba,Hawk2}
is not clear, since the semiclassical approach used there is not
justified for two-dimensional gravity which has a dimensionless
coupling \cite{Polch}.
However,
the two-dimensional target space models offer
hope that non-perturbative effects may guarantee $\Lambda=0$
even beyond the semiclassical approximation, which
has in any case never been derived even in the four-dimensional case.
The approach of refs. \cite{Col,Ba,Hawk2} required some hierarchy
between the Planck scale ($M_P$)
and the sizes of the wormholes considered,
and would `iron out' the cosmological constant only if it was
`pre-cooked'
to be small compared with $M_P^4$, as in generic
models with approximate supersymmetry and no-scale supergravity
models in particular \cite{Sup}.
Thus our ideas are not obviously
contradictory with those of refs. \cite{Col,Ba,Hawk2}.
\paragraph{}
{
{\Large{\bf Acknowledgements}} \\

We thank C. Bachas, I. Bakas, E. Kiritsis,
N.D. Vlachos and S. Yankielowicz for discussions. The
work of D.V.N. is partially supported by DOE grant
DE-FG05-91-ER-40633.


\newpage
\begin{thebibliography}{99}
\bibitem{Ell} J. Ellis, N.E. Mavromatos and D.V. Nanopoulos,
CERN-TH.6147/91, CTP-TAMU-41/91, to appear in Phys. Lett. B.
\bibitem{Gro} D.J. Gross and A.A. Migdal, Phys. Rev. Lett. 64 (1990),
717;
\par M. Douglas and S. Shenker, Nucl. Phys. B335 (1990), 635;
\par E. Brezin and V. Kazakov, Phys. Lett. 236B (1990), 144.
\bibitem{Ant} I. Antoniadis, C. Bachas, J. Ellis and
D.V. Nanopoulos, Phys. Lett. B211 (1988), 393.
\bibitem{Witt} E. Witten, Phys. Rev. D44 (1991), 314.
\bibitem{Kleb} A. Sengupta
and S. Wadia, Int. Jour. Mod. Phys. A6 (1991), 1961;
\par D.J. Gross and I.R. Klebanov, Nucl. Phys.
B352 (1991), 671.
\bibitem{Jev} S.R. Das and A. Jevicki, Mod. Phys. Lett. A5 (1990),
1639.
\bibitem{Ava} J. Avan and A. Jevicki, Phys. Lett. B266 (1991), 35.
\bibitem{Pop} I. Bakas and E. Kiritsis,
Nucl. Phys. B343 (1990), 185;
Progr. Theor. Phys. Suppl. 102 (1990), 15.
\par C.N. Pope, L.J. Romans and X. Shen,
Phys. Lett. B236 (1990), 173; Nucl. Phys. B339 (1990), 191.
\bibitem{Bak} I. Bakas, Phys. Lett. B228 (1989), 57;
\par C.N. Pope, L.J. Romans and X. Shen, {\it A brief history
of $W_{\infty}$}, in {\it Strings '90}, World Scientific (1991).
\bibitem{Hag} J. Ellis, J.S. Hagelin,
D.V. Nanopoulos and M. Srednicki, Nucl. Phys. B241 (1984), 381.
\bibitem{Gross} D.J. Gross, Nucl. Phys. B236 (1984), 349.
\bibitem{Lahi} T.J. Allen, M.J. Bowick and
 A. Lahiri, Phys Lett.
237B (1990), 47.
\bibitem{Hawk} S. Hawking, Comm. Math. Phys. 43 (1975), 199;
{\it ibid} 87 (1982), 395.
\bibitem{Pol} A.M. Polyakov, Mod. Phys. Lett. A6 (1991), 635.
\bibitem{MS} G. Moore and N. Seiberg, Rutgers University preprint
RU-91-29,
YCTP-P19-91
(1991).
\bibitem{MS2} G. Moore, N. Seiberg and M. Staudacher,
Rutgers/Yale preprint RU-91-11, YCTP-P11-91, Nucl. Phys. B in press.
\bibitem{Ven} G. Veneziano, Phys. Lett. B167 (1986), 388.
\par J. Maharana and G. Veneziano, Nucl. Phys. B283 (1987), 126.
\bibitem{Zam} A.B. Zamolodchikov, Sov. J. Nucl. Phys. 44 (1987), 529;
{\it ibid} 46 (1987), 1090.
\bibitem{Ovr} M. Evans and B. Ovrut, Phys. Rev. D39 (1989), 3016;
{\it ibid} D41 (1990), 3149.
\bibitem{Mav} G. Veneziano, unpublished;
\par J. Maharana, N.E. Mavromatos and
G. Veneziano, unpublished.
\bibitem{W-grav} E. Bergshoeff, C.N. Pope, L.J. Romans,
E. Sezgin and X. Shen, Mod. Phys. Lett. A5 (1990) 1957;
\par See also Pope {\it et al.} in ref. \cite{Bak}.
\bibitem{Shen} X. Shen and X.J. Wang, Texas AM-University
preprint CTP TAMU-59/91 (1991).
\bibitem{Witten} D. Gross, Phys. Rev. Lett. 60 (1988), 1229.;
\par E. Witten, Institute for Advanced Studies preprint,
IASSNS-HEP-88/55, based on talk given in the
Royal Society meeting on {\it String Theory}, London,
December 1988.
\bibitem{Com} S. Coleman, Comm. Math. Phys. 31 (1973), 259.
\bibitem{Mand} G. Mandal, A. Sengupta and S. Wadia, Institute
of Advanced Studies preprint, Princeton IASSNS-10-91 (1991).
\bibitem{Witt2} E. Witten, Institute for Advanced Studies preprint,
IASSNS-HEP-91/51.
\bibitem{Lian} B. Lian and G. Zuckerman, Phys. Lett. B254 (1991), 417;
{\it ibid} B266 (1991), 21.
\bibitem{Sak} A. Jevicki and B. Sakita, Nucl. Phys.
B165 (1980), 511; {\it ibid} B185 (1981), 89.
\bibitem{Polch2} J. Polchinski, Texas-Austin University preprint
UTTG-91-06 (1991), Nucl. Phys. B in press.
\bibitem{Min}D. Minic and Z. Yang, Austin-Texas University preprint
UTTG-23-91 (1991).
\bibitem{Tri} D.J. Gross and I. Klebanov, Nucl. Phys.
B359 (1991), 3.
\bibitem{Gor} V. Gorini, A. Frigerio, M. Verri, A. Kossakowski and
E.C. Sudarshan, Rep. Math. Phys. 13 (1978), 149.
\bibitem{Park} Q-Han Park, Phys. Lett. B236 (1989), 429; {\it ibid}
B238 (1990), 287; see also E. Witten, in {\it Strings '90},
World Scientific (1991).
\bibitem{Sup} E. Cremmer, S. Ferrara,
C. Kounnas and D.V. Nanopoulos,
Phys. Lett. B133 (1983), 61;
\par J. Ellis, C. Kounnas and D.V. Nanopoulos,
Nucl. Phys. B241 (1984), 406 and
Nucl. Phys. B247 (1984), 373.
\bibitem{Sustri} E. Witten, Phys. Lett. B155 (1985), 151.
\bibitem{Col} S. Coleman, Nucl. Phys. B307 (1988), 867; {\it ibid}
B310 (1988), 643.
\bibitem{Ba} E.B. Baum, On topology in Euclidean Gravity, Princeton
University preprint, 82-0504 (1985).
\bibitem{Hawk2} S. Hawking, Phys. Lett. B195 (1987), 337; Nucl.
Phys. B335 (1990), 155.
\bibitem{Polch} J. Polchinski, Nucl. Phys. B324 (1989), 123.
\end{thebibliography}
\end{document}